# Dynamics of the dissociative electron attachment to Ethanol

Sukanta Das, Suvasis Swain, and Vaibhav S. Prabhudesai

Tata Institute of Fundamental Research, Colaba, Mumbai 400005 India

*vaibhav@tifr.res.in

**Abstract:**

We report the detailed dynamics of the site selectivity observed in the dissociative electron attachment (*DEA*) process in ethanol based on the momentum images obtained using the velocity slice imaging technique. The H⁻ dissociation channel shows the site selectivity where the anion signal from the O-H site peaks at 6.5 eV and 8 eV, and that from the C-H site peaks at 9.5 eV. The momentum images also show the two-body dissociation dynamics for the O-H site break-up. This dissociation channel shows a substantial effect of the torsion mode of vibrations on the electron attachment process. In contrast, the C-H site dissociation results from the many-body break-up consistent with the earlier reports of *DEA* dynamics from organic molecules. We have also found that the OH⁻ channel has a resonance at 9.3eV and is produced with very little kinetic energy. Using the isotope substitution, we show the role of H atom scrambling in the C-O bond dissociation leading to the OH⁻ channel. This channel shows a substantial deviation from the corresponding photodissociation dynamics.

## I. Introduction:

The study of the interaction of low-energy electrons with molecules has great importance because when high-energy radiation like x-rays, gamma rays, or energetic charged particles interact with matter, they produce secondary low-energy electrons. These low-energy electrons contribute significantly to DNA damage [1]. Dissociative electron attachment (*DEA*) plays a vital role in this process. In a bottom-up approach to understanding the details of the underneath complex dynamics of DNA strand breaks, studies of *DEA* to several simple organic molecules like carboxylic acids, alcohols, and simple aromatic compounds have been taken up as a starting point. On the other hand, *DEA* shows the site selectivity in organic molecules [2]. The site-selectivity observed in the *DEA* process has been for electron energies well beyond the respective sites' dissociation threshold. This points to the vast potential of low-energy electron-based chemical control. To realise this feature, understanding the underlying dynamics of the *DEA* process that causes site selectivity is of utmost importance. Besides, the excited states of negative ions play a vital role in various applications like electron beam lithography and plasma processing [3]. However, details of these dynamics are complicated to obtain from the theoretical calculations due to the resonant nature of the underlying excited negative ion states.

Simple alcohols are used to identify the site-selectivity for the O-H bond in the organic molecules [4]. It has been pointed out that the dissociation of the O-H bond in the alcohols shows a direct resemblance in the DEA peaks in the H⁻ ion yield curve with that observed in the water, which is the precursor molecule for the hydroxyl group. Here, we have carried out detailed measurements of *DEA* to ethanol ($C_2H_5OH$) to unravel the dynamics of this process in the case of simple alcohols.

Earlier works on DEA measurement on ethanol are carried out by Prabhudesai *et al.* [2, 4], Ibanescu and Allan [5], Orzal *et al.* [6] and Wang *et al.* [7]. Three peaks in the H⁻ channel at 6.4eV, 7.8eV, and 9.3 eV were first reported by Prabhudesai *et al*. [2]. They also reported the absolute cross-section of these resonances [4]. Ibanescu and Allan [5] introduced partially deuterated ethanol ($C_2H_5OD$) and reported that 6.35eV and 7.85eV resonances are due to H⁻ from the OH site, and 9.18eV resonance is from the CH site. Besides, they have also obtained the $C_2H_5^-$ ions peaking at 2.75eV, 6.35eV, and 9.15eV. They have reported the vibration excitation cross-section and photoelectron spectra (PES) and compared *DEA* spectra with those measurements. Orzal *et al.* [6] have reported O⁻ (5.8eV), OH⁻ (8.2eV), and $C_2H_5^-$ (2eV, 5eV, and 8.2eV) and discussed the possible channels of these anions. Wang *et al.* [7] have done the momentum imaging of O⁻/OH⁻ at 6eV and 9eV using Velocity Slice Imaging (*VSI*). Due to the low energy and mass resolution of the *VSI* spectrometer, they were unable to separate O⁻ and OH⁻.

Although the strong site selectivity in *DEA* is seen in the H⁻ channel, which is also the most abundant channel, there has been no report of the dynamics resulting in this process. We have measured these ions' angular distribution and kinetic energy distribution at various electron energies to unravel the dissociation dynamics of the temporary negative ion formed by the electron attachment. To identify the site-selective signal of the H⁻ ions, we have carried out the measurements on partially deuterated ethanol ($CH_3CH_2OD$) and obtained the momentum images for H⁻ and D⁻ separately. We also looked for other negative ions and have identified the OH⁻ ions by improving the mass resolution of the spectrometer. This ion signal peaks around 9.3eV. In this paper, we report the details of these findings and interpret the underlying dynamics using the kinetic energy and angular distributions.

## II. Experiment

The details of the experimental setup are given elsewhere earlier [8], and here we describe the experiment in brief. The experiment is carried out in a crossed electron beam–molecular beam geometry where the effusive molecular beam produced by the capillary array is put along the axis of the velocity slice imaging (*VSI*) spectrometer. The spectrometer (Figure 1) consists of a pusher, puller, four electrostatic lens electrodes, a Flight tube, and a 2-D position-sensitive detector comprised of a Micro Channel Plate (*MCP*) detector, a Phosphor screen, and a CCD camera.

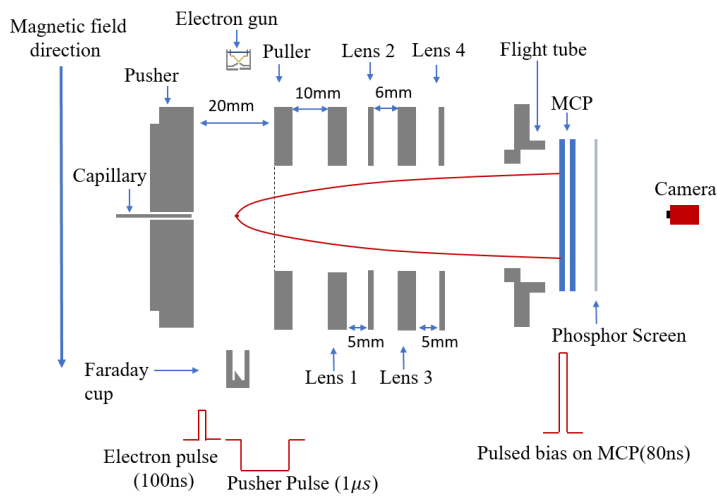

Figure 1. Schematic of the velocity slice imaging spectrometer used in the experimental setup

The interaction volume, which is due to the overlap of the electron beam and the molecular beam, is at the centre of the region between the pusher and puller. The magnetically collimated pulsed electron beam is produced from a home-built electron gun. The electron beam collimating magnetic field of 50 Gauss strength is produced by an externally mounted pair of magnet coils in the Helmholtz geometry co-axial to the electron gun. A Faraday Cup, co-axial to the electron gun on the opposite side of the interaction zone, is used to measure the current. We apply a delayed (100 ns with respect to the electron pulse) negative voltage pulse on the pusher of height 80V and 1μs width to extract the ions from the interaction zone. Outside the chamber, a Baratron is connected to the back of the capillary to measure the pressure behind the capillary. The voltage on the detector assembly is pulsed using a variable width high voltage switch. The detector pulsing is synchronized with the central part of the ion time of flight peak to obtain the appropriate slice of the Newton sphere. The width of the detector pulse is kept at 80 ns (the minimum possible from such a switch) for slice imaging.

The whole chamber is kept below $5 \times 10^{-7}$ torr base pressure during the experiment. The gas line used to introduce the sample into the chamber and the needle valve used to maintain the gas flow are heated up to a constant temperature of $40^0$ C throughout the experiment. Measurements are carried out in two steps. The ion yield curve is obtained in the first step, and its resonance position is calibrated using O⁻ from $O_2$ [9]. In the second step, The pixel image of the illuminated Phosphor screen is recorded by the CCD camera. These pixel images are converted to momentum images by calibrating against H⁻ from $H_2$. From these momentum images, kinetic energy (*KE*) distributions and angular distributions are obtained for the detected anions.

### III. Results and Discussion

We have observed three resonances for H⁻ from ethanol at 6.5eV, 8eV, and 9.5eV, as shown in Figure 2(a), consistent with the previously reported results [2, 5]. Using partially deuterated ethanol (Figure 2), we have identified that H⁻ from the O-H site is responsible for the 6.5eV and 8 eV peaks and H⁻ from $C_2H_5$ peaks at 9.5 eV consistent with the earlier report [5].

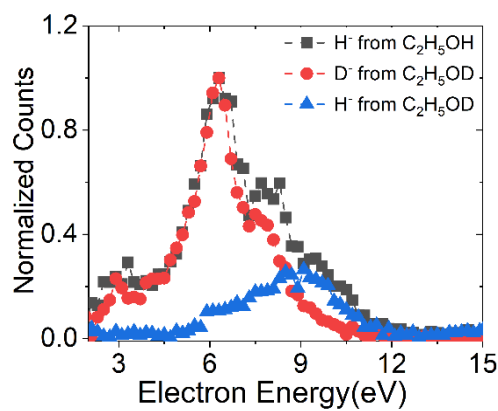

Figure 2. Comparison of the normalised ion yield curve for H⁻ from ethanol (black square) and D⁻ (red circles) and H⁻ (blue triangles) from partially deuterated ethanol ($C_2H_5OD$).

We obtained the momentum images of H⁻ from $C_2H_5OH$ at 6.5, 8 and 9.5eV and D⁻ at 6.5 and 8eV and H⁻ at 9.5 eV from $C_2H_5OD$ (Figure 3). As discussed in the earlier work, the momentum images were obtained for the crossed beam and static gas that contributes to the background signal [8]. The images shown in Figure3 are obtained after subtracting the static gas images from the crossed beam images with appropriate normalization. Due to the presence of the transverse magnetic field, the H⁻ ions, which are the lightest, follow the deviated trajectories. This shifts the momentum image to one side of the spectrometer axis and introduces distortion. This is evident in Figure 3. For the image analysis, we have considered only half of the image obtained close to the centre of the detector, which has the least distortion, as discussed in ref. 8. In momentum images, we observe that at 6.5 eV, H⁻ is mainly scattered in the backward direction (Figure 3(a)). As can be seen from the images obtained from the partially deuterated sample, the site selectivity of the hydride ion signal in the ion yield curve is also evident in the momentum images. The 6.5eV image of both H⁻ from $C_2H_5OH$ and that for D⁻ from $C_2H_5OD$ shows similar momentum images (Figure 3 (a) and (d)). As the 9.5eV peak has a long tail which contributes to the 8eV image, one can see a clear signature of the low energy blob in the momentum image of H⁻ at 8eV from $C_2H_5OH$ (Figure 3(b)), which is absent in the image of D⁻ from $C_2H_5OD$ (Figure 3(e)) whereas the blob is present in H⁻ image at 9.5eV from both $C_2H_5OH$ and $C_2H_5OD$ (Figure 3 (c) and (f)). The *KE* distribution obtained from the momentum images of the H⁻ ions is shown in Figure 4.

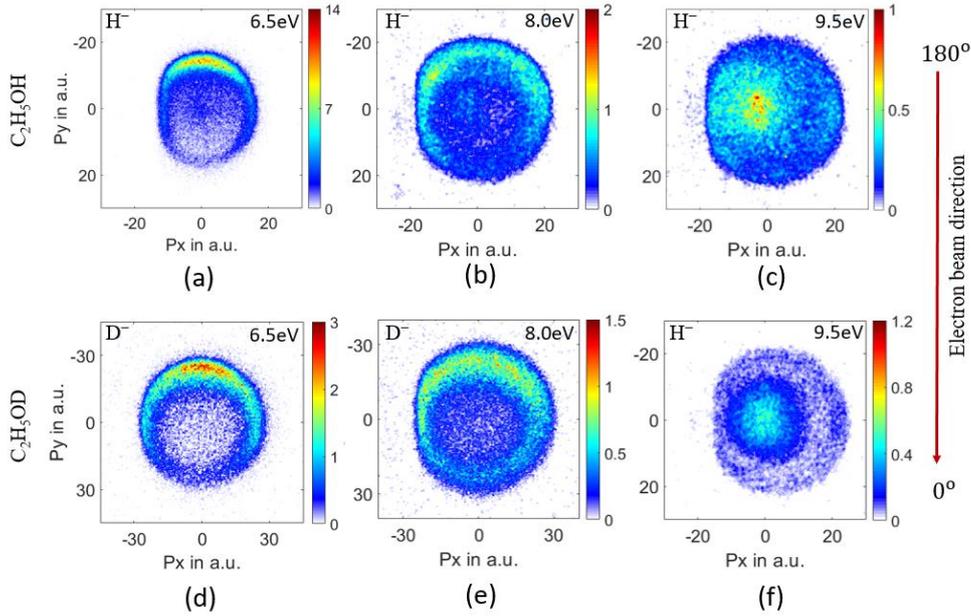

Figure 3. Momentum images of H⁻ from $C_2H_5OH$ at (a) 6.5eV (b) 8eV and (c) 9.5eV electron energies. Momentum images of D⁻ at (d) 6.5eV, (e) 8eV, and H⁻ at (f) 9.5eV from $C_2H_5OD$. The arrow indicates the direction of the electron beam.

O'Malley and Taylor [10] reported the detailed theoretical calculation for the angular distribution of the negative ion produced in *DEA* to diatomic molecules under the assumption that (i) there is only one resonance state that contributes to the negative ion formation, (ii) the negative ion state does not rotate before it decays (iii) the coupling is spin-independent. Azria *et al.* [11] adopted a similar treatment for polyatomic molecules. They obtained the expression for the angular distribution of the fragment anions as

$$I(\theta) \propto \frac{1}{2\pi} \int_0^{2\pi} \left| \sum_{l,m,\gamma} i^l \exp(i\delta_l)\, a_{lm}^{\gamma} X_{lm}^{\gamma*}(\theta,\varphi) \right|^2 d\varphi \qquad (1)$$

where $X_{lm}^{\gamma*}(\theta,\varphi)$ is the basis function for the irreducible representation of the group of molecules, $a_{lm}^{\gamma}$ their amplitude and $\delta_l$ their phase. Here the angles $(\theta,\varphi)$ determine the orientation of the dissociating bond about the incoming electron beam. The above functions are in the dissociation frame of the molecule and expressed as a linear combination of spherical harmonics with appropriate frame transformation from the lab frame to the molecular frame.

Ethanol belongs to the $C_s$ molecular point group, with only two irreducible representations, namely $A'$ and $A''$. In the ground state, 26 electrons of this molecule are arranged in 13 doubly occupied orbitals. The ground-state electron configuration is written as [12]

$(1a')^2 (2a')^2 (3a')^2 (4a')^2 (5a')^2 (6a')^2 (7a')^2 (1a'')^2 (8a')^2 (9a')^2 (2a'')^2 (10a')^2 (3a'')^2$

which corresponds to an $A'$ state. The closest unoccupied molecular orbitals (*MO*s) are $(11a')$, $(12a')$, and $(4a'')$. The expected angular distribution under the axial recoil approximation for the $A'$ to $A'$ transition using the first two partial waves would be

$$I_{s+p} = \frac{a^2}{2} + \frac{b^2}{2}(cos^2\ \theta) + ab\ cos\ \theta\ cos\delta_1 \qquad (2)$$

where $a$, and $b$ indicate the relative contributions of the two partial waves and $\delta_1$ is the relative phase between the $s$ and $p$ waves, respectively. For $A'$ to $A''$ transition using the first two allowed partial waves ($p$ and $d$) would be

$$I_{p+d} = \frac{a^2}{2}(sin^2\ \theta) + b^2\left(\frac{3}{8}\ (sin2\theta)^2\right) + \frac{\sqrt{3}}{2}ab\ cos\delta_1 sin\theta\ sin\ 2\theta \qquad (3)$$

We have obtained the angular distribution of the H⁻ ion from the momentum images and analysed them using combination of equation (2) and equation (3) under the axial recoil approximation.

Here we discuss the *DEA* dynamics that result in H⁻ ion formation at three different resonances. The dynamics of OH⁻ ion formation is addressed in the end.

### A. H⁻ formation at the first resonance

As concluded from the ion yield curves of the partially deuterated sample, for this 6.5eV resonance, the H⁻ is formed from the -OH site. The most obvious path for the H⁻ formation is the direct cleavage of the O-H bond according to the reaction

$$e + C_2H_5OH \rightarrow (C_2H_5OH^-)^* \rightarrow C_2H_5O + H^- \qquad (4)$$

where $C_2H_5OH^-*$ represents the temporary negative ion (*TNI*) state at 6.5 eV. Using established heat of formation values of ethanol (-234kJ/mol), $C_2H_5O$ (17kJ/mol) and H (218kJ/mol) and subtracting the electron affinity of H (72.8kJ/mol), we get 396kJ/mol or 4.1eV as the minimum energy required for the formation of H⁻ [13]. Since this channel is a two-body breakup, the excess energy of 2.4 eV would be distributed among the two fragments inversely proportional to their masses. Accordingly, up to 2.34 eV of energy would appear as the *KE* of the H⁻ ion. The observed *KE* distribution of the H⁻ ions (Figure 4(a)) extends up to 2.75eV, peaking around 2eV. The maximum *KE* observed in this channel is consistent with the threshold of the above channel, considering the spread in the electron energy, which is about 0.8eV. The spread in the *KE* distribution results from this energy uncertainty, momentum imaging resolution, which is about 0.3eV and the internal excitation of the neutral $C_2H_5O$ fragment.

The angular distribution obtained for this channel is shown in Figure5(a). For the $A'$ to $A'$ transition, all partial waves can contribute starting from $l=0$, whereas, for the $A'$ to $A''$ transition, partial waves with $l \geq 1$ would contribute to the electron capture. The significant difference between these two transitions is that for the $A'$ to $A''$ transition, the angular distribution of the ions formed from the cleavage of the bond lying in the symmetry plane of the molecule would always have nodes at $0^0$ and $180^0$ about the electron beam. On the other hand, for the $A'$ to $A'$ transition due to the contribution from the *s*-wave, the ion signal would have finite strength in the $0^0$ and $180^0$ about the electron beam. This is evident from the angular distributions obtained for the 6.5eV and 8.5eV channels in water [14,15]. The 6.5eV resonance is understood as the one specific to the O-H site; it is expected to show the angular distribution consistent with that obtained for water. In water, the resonance responsible for this peak corresponds to the $B_1$ state, a valance excited resonance associated with the excitation of the lone pair of electrons from the O atom. Ibanescu and Allan [5] have reported that the 6.5eV resonance of H⁻ is the Feshbach type formed by the excitation of the electron from 3a″ orbital (comprising mainly of the lone pair of electrons on the O atom) to the higher empty MOs while the electron capture. This transition is the basis of the O-H site selectivity observed in the simple alcohols [2].

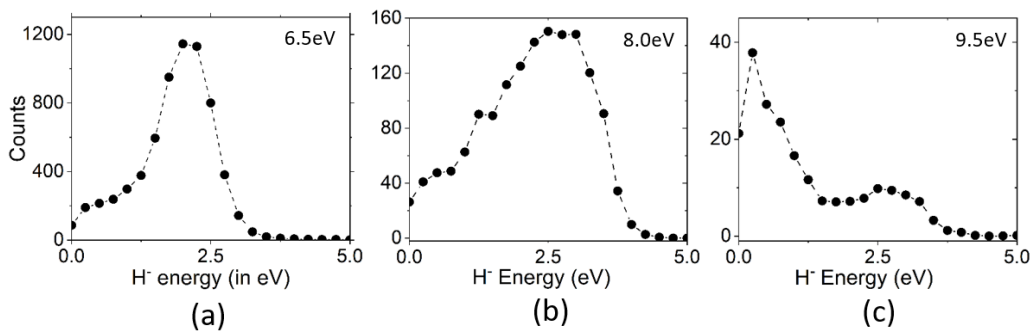

Figure 4. Kinetic energy distribution of H⁻ from $C_2H_5OH$ at (a) 6.5eV, (b) 8eV and (c) 9.5eV obtained from momentum images.

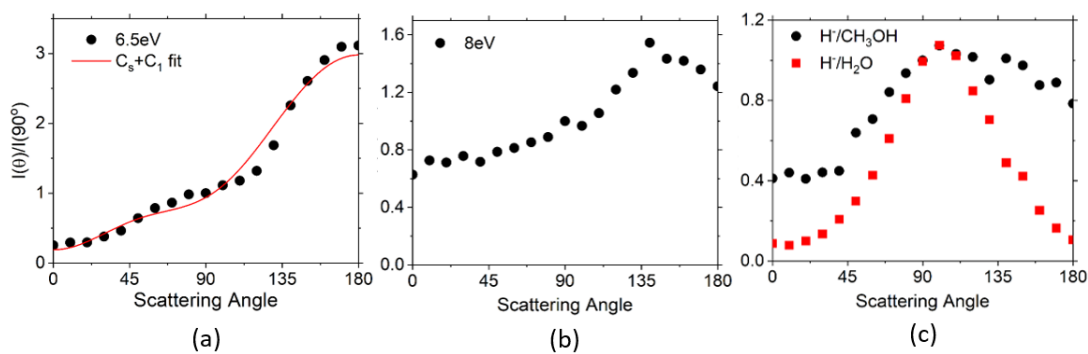

Figure 5. Angular distribution of H⁻ ions obtained at (a) 6.5eV , and (b) 8eV electron energy from ethanol and (c) at 6.5eV from methanol (black circle) and water (red square). The solid curves show the fit obtained by combining equation (2) and (3)

To further understand the DEA dynamics, we compare the angular distribution of H⁻ ion from ethanol obtained at 6.5eV with that obtained for water at 6.5eV. As shown earlier, for water, the angular distribution of the *DEA* signal peaks around $100^0$ with no intensity in the $0^0$ and $180^0$ directions (Figure 5(c)) [14]. The site selectivity of the O-H bond cleavge in DEA is understood as the result of a 2-particle-1-hole resoancne that is formed by excitation of the lone pair of electrons from the O atom. This is also seen in water molecule. With the consideration of similar underlying resonance, we expect similar angular distribution at this resonance from alcohols as the molecules retain the reflection symmetry about the plane containing the O-H bond. For example, for a molecule like simple alcohol that belongs to the $C_s$ symmetry group, this resonance would be of $A''$ character and would not have any signal in the $0^0$ and $180^0$ angles under the axial recoil approximation. However, the angular distribution obtained from the 6.5eV resonance image for ethanol shows no nodes at $0^0$ and $180^0$.

Table I: Fitting parameters in equation (5) obtained for the angular distribution of H$^-$ions at 6.5eV.

| Energy | $C_1$ (s + p partial waves) | | | $C_s$ (p + d partial waves) | | | $C_1 : C_s$ |
|---|---|---|---|---|---|---|---|
| | $a_1$ | $b_1$ | $\delta_1$ | $a_2$ | $b_2$ | $\delta_2$ | |
| 6.5 eV | 1.17 | 1.34 | 2.63 | 0.82 | 0.58 | 6.21 | 1.0 : 0.75 |

Similar results were also obtained for methanol by Slaughter *et. al* [16]. In their work, they showed the entrance channel amplitude of for the corresponding resonance at 6.5eV having no intensity in the plane containing the O-H bond. This is consistent with our understanding of the structure of the underlying resonance. However, the angular distribution of H$^-$ ion from methanol shows no node around $0^0$ and $180^0$ direction about the electron beam (Figure 5(c)). They explained this observation as the loss of $C_s$ symmetry due to torsional vibrations in the molecule about the C-O bond and opening of COH angle. As can be seen from the Figure5(a), in the case of ethanol, the relative intensity of the H$^-$ ion signal in the $180^o$ direction w.r.t $90^o$ is even higher than that observed in methanol. We also give the corresponding angular distribution data for methanol for the relevant resonance in Figure 5(c).

For ethanol, the vibrational energy of torsional modes that involve the out-of-plane motion of O-H bond are 200 and 251cm$^{-1}$ [17, 18] and for methanol it is around 295cm$^{-1}$ [13]. These torsional motions break the reflection symmetry of the molecule. Under our experimental condition this leads to a population of around 35% in the torsionally active states for Ethanol, and around 24% for methanol. If we assume torsionally ground and excited molecules participating in the dissociation under axial recoil approximation then for torsionally ground sate molecule, $C_s$ symmetry holds, hence can taken $A'$ to $A''$ transition (equation 3). For the molecules which are torsionally excited, we can take $C_1$ symmetry (equation 2). We have used the incoherent sum of the angular distribution functions for the torsionally ground state molecules (equation3) and torsionally excited molecules (equation 2) tp fit the observed angular distribution. The used fitting function is

$$I_{C_1+C_S} = C_1 \left( \frac{a_1^2}{2} + \frac{b_1^2}{2}(cos^2\theta) + a_1 b_1 cos\theta cos\delta_1 \right) + C_s \left( \frac{a_2^2}{2}(sin^2\theta) + b_2^2 \left( \frac{3}{8}(sin2\theta)^2 \right) + \frac{\sqrt{3}}{2} a_2 b_2 cos\delta_2 sin\theta sin 2\theta \right) \quad (5)$$

where, $C_1$ and $C_S$ are the corresponding co-efficients for the two contributing sets of molecules.

According to the fit (Table I), almost 57% contribution is found to be from the $C_1$ symmetry molecules and the rest from the $C_s$ symmetry molecules. The observed contribution is much larger than the that expected from the vibrational excitation. Another reason for such a break of reflection symmetry could be due to the presence of two conformers in the ground state of ethanol, namely anti ($C_s$ symmetric) and gauche ($C_1$ symmetric). Most of the microwave and IR studies have reported anti is the most stable one, but there is only a 40cm$^{-1}$ energy gap between the gauche and anti-conformer [19, 20], hence a predominance is expected of the gauche conformer at room temperature due to its two-fold degeneracy. Barnes and Hallam [21] estimated the ratio of anti and gauche conformer at room temperature in the vapour phase to be around 2:1. Shaw *et. al.*[22] reported it to be 42:58 from their IR studies.

Based on these observations, we conclude that the large signal obtained in the backward direction for the H$^-$ ion signal is due to the combination of torsional excitation available in the molecule at room temperature and the asymmetric conformers present in the target beam. It can also be possible that for such torasionally excited molecules, the autodetachment cross-section may also vary depending on the change in geometry affecting the DEA signal strength.

## B. H$^-$ formation at the second resonance

The second resonance is also from the -OH site. If we consider the same channel as the first resonance i.e.

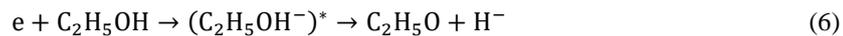
$$e + C_2H_5OH \rightarrow (C_2H_5OH^-)^* \rightarrow C_2H_5O + H^- \qquad (6)$$

then the excess energy will be 3.9eV. If we see the images of 6.5eV and 8eV resonance, they show different angular distributions. The obtained distribution is shown in Figure 5(b). For the 6.5eV resonance, the ion intensity peaks in the backward direction (close to 180°), whereas for the 8eV resonance, it peaks around 135°. This observation and the observed *KE* distributions may indicate that the *DEA* dynamics of this resonance involve substantial internal excitation before dissociation. A comparison of *KE* distribution at these two energies (Figure 4(a) and (b)) shows that for 8 eV resonance, the neutral fragment is formed with a broader distribution of internal energy. This is possible if the dissociation dynamics consist of slow dissociation to begin with before reaching the final slope. This can result in substantial distortion in the molecule, which can result in the excitation of various vibrational modes of the anion. Similar dynamics have also been reported for water for its 8.5eV resonance [23].

Another possible channel is:

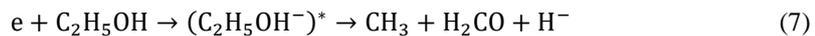
$$e + C_2H_5OH \rightarrow (C_2H_5OH^-)^* \rightarrow CH_3 + H_2CO + H^- \qquad (7)$$

where the TNI undergoes a three-body dissociation. Using the heat of the formation of CH$_3$ (145.69kJ/mol) and H$_2$CO (-115.9kJ/mol), we obtain the thermodynamic threshold for this channel as 4.23eV [13]. In that case, the system has 3.77eV of excess energy. We cannot rule out this channel based on the kinetic energy distribution.

## C. H$^-$ formation at the third resonance

For the third resonance, based on the site selectivity observed in the ion yield curve of the partially deuterated molecule, H⁻ can be formed by direct bond cleavage from either the $CH_3$ site or $CH_2$ site, and the possible channels can be

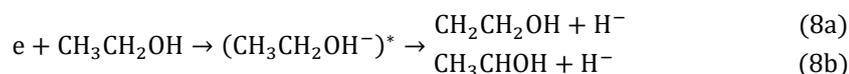

$$e + CH_3CH_2OH \rightarrow (CH_3CH_2OH^-)^* \rightarrow \begin{array}{ll} CH_2CH_2OH + H^- & \quad (8a) \\ CH_3CHOH + H^- & \quad (8b) \end{array}$$

Using the heat of formation for $CH_2CH_2OH$ (-25.9kJ/mol) and $CH_3CHOH$ (-54kJ/mol) [24], the thermodynamic threshold for the first channel is 3.66eV and the second channel is 3.37eV. But the resonance occurs around 9eV, so more than 5.5eV excess energy is available in the system. However, the *KE* obtained in the H⁻ channel is less than 1eV. This implies that more than 4.5eV energy is excess which can appear as the internal energy of the fragments.

This energy is sufficient to break the molecular fragment further into smaller fragments. This can also result from a three or multiple body dissociation, which has several possibilities like

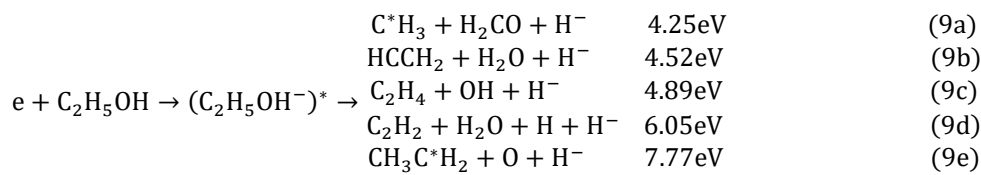

$$e + C_2H_5OH \rightarrow (C_2H_5OH^-)^* \rightarrow \begin{array}{lll} C^*H_3 + H_2CO + H^- & 4.25eV & (9a) \\ HCCH_2 + H_2O + H^- & 4.52eV & (9b) \\ C_2H_4 + OH + H^- & 4.89eV & (9c) \\ C_2H_2 + H_2O + H + H^- & 6.05eV & (9d) \\ CH_3C^*H_2 + O + H^- & 7.77eV & (9e) \end{array}$$

In all these channels except the last two channels where H⁻ appears to arise from the OH site, can contributing to this resonance. Hence we conclude that at 9.5eV, the C-H bond cleavage is associated with the multiple fragmentation process.

### D. OH⁻ formation at 9.3eV

We also looked for the heavier ions but only detected OH⁻ which shows a peak in the ion yield curve around 9.3eV (Figure 6(a)). In the photodissociation study of methanol, the presence of high energy OH radicle at 157nm (~7.89eV) was reported by Yang *et. al* [25]. According to *ab initio* calculations, there is an avoided crossing between 3p and 3s surface. Oxygen loan pairs are excited to 3p state by 157nm photon, by internal conversion they transfer to 3s state which leads to breaking of C-O bond. The presence of high energy OH in this photodissociation channel shows it as a two-body breakup. Since oxygen loan pairs are involved in the process, a similar dissociation of OH in ethanol can be expected [26]. In our case, we detected OH⁻ which shows a peak in the ion yield curve at around 9.3eV. The lowest energy pathway for OH⁻ formation is the cleavage of the C-O bond in a two-body breakup.

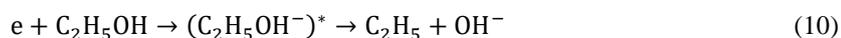

$$e + C_2H_5OH \rightarrow (C_2H_5OH^-)^* \rightarrow C_2H_5 + OH^- \qquad (10)$$

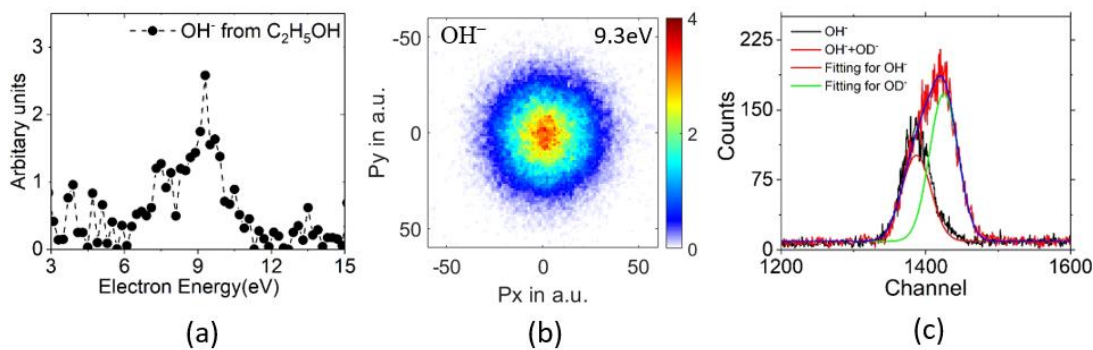

Figure 6. (a) Ion yield curve for OH⁻ from ethanol (b) Momentum image of OH⁻ obtained at 9.3eV from ethanol. (c) A part of the mass spectrum obtained for ethanol and partially deuterated ethanol ($C_2H_5OD$) at 9.3 eV.

Using the heat of formation of $C_2H_5$ (119kJ/mol) and electron affinity of OH (176.34kJ/mol), we arrive at the value of 2.24eV as the thermodynamic threshold for OH⁻ formation from this channel [13]. But we have observed the ion yield peak at 9.3eV, which is 7eV above the thermodynamic threshold and 1.4eV above the photodissociation limit. We have also found both OH⁻ and OD⁻ from $C_2H_5OD$, as we can see from the mass spectrum (Figure 6(c)), where we have fitted the two peaks (OH⁻ and OD⁻) with Gaussians of appropriate widths. These peaks indicate that the hydrogen scrambling occurs before dissociation. VSI image (Figure 6(b)) of OH⁻ also shows very low kinetic energy (0.5eV) in this ion. Hence, we can expect that more than one fragment may be formed during the DEA process. One possibility of such channels is

$$e + C_2H_5OD \rightarrow (C_2H_5OD^-)^* \rightarrow \begin{array}{ll} C_2H_4 + OD^- + H & (11a) \\ C_2H_4 + OH^- + D & (11b) \\ C_2H_3D + OH^- + H & (11c) \end{array}$$

However, as one of the fragments is an H/D atom, it would carry most of the excess energy, leaving very little kinetic energy in the anion channel. This also suggests that the parent state of the resonance responsible for OH⁻ formation is different from the photodissociation channel. The obtained momentum image of OH⁻ is consistent with this picture (Figure 6(b)).

## IV. Conclusion

We have shown that *DEA* to ethanol shows a strong site selectivity, i.e. O-H and C-H sites follow their characteristic dissociation pattern. We have measured the kinetic energy and angular distribution of the H⁻ channel and found that the O-H site breakage results from a two-body dissociation without any scrambling. The 6.5eV resonance shows a substantial effect of the torsion mode of vibrations in the electron attachment process, which manifests in the observed angular distribution of the H⁻ ions. The gauche conformer of the molecule may also be responsible for the observed deviation of the angular distribution from that expected under axial recoil approximation. The 8eV resonance shows considerable energy left in the internal excitation of the molecular fragment. In contrast, the C-H site breaking corresponding to the 9.5eV resonance is associated with the many-body breakup consistent with the earlier reports for C-H sites in other organic molecules. We have also measured the OH⁻ channel in terms of its kinetic energy and angular distribution and conclude that this channel is associated with hydrogen scrambling and many-body breakup.

**Acknowledgement**

The authors acknowledge the financial support from the Department of Atomic Energy, India, under Project Identification No. RTI4002.